\input{aipcheck}

\documentclass[
    ,final            
  ]
  {aipproc}

\layoutstyle{8x11double}

\begin{document}

\title{\rightline{\normalsize\rm IIT-CAPP-11-01}\vspace{.05in}From Neutrino Factory
to Muon Collider}


\author{Daniel M. Kaplan}{
  address={Illinois Institute of Technology, Chicago, IL 60616, USA}
}

\begin{abstract}
After summarizing the important commonalities between neutrino factories and muon colliders, the key differences are discussed. These include a much larger needed cooling factor ($\sim$\,10$^6$ in six-dimensional emittance), a smaller number of muon bunches (perhaps only one of each charge), and acceleration to much higher energy, implying significantly different technical choices for some of the cooling and acceleration subsystems. The final storage rings are also quite different. Nevertheless, a neutrino factory could serve as a key stepping stone on the path to a muon collider.
\end{abstract}

\maketitle


\section{Introduction}
Although designs without cooling have been discussed, a stored-muon-beam neutrino factory benefits from some muon-beam cooling. Recent designs feature a factor $\approx$\,3 reduction in transverse emittance, accomplished using an $\approx$\,80-meter-long 
transverse cooling channel~\cite{Neuffer-NuFact10,JINST,IDS}.
A muon collider, on the other hand, requires substantial muon cooling in order to reach the physically interesting luminosity regime, ${\cal L}\sim10^{30}$--10$^{34}$\,cm$^{-2}$s$^{-1}$. The former luminosity figure suffices for studies of a possible $Z^\prime$ or technihadron state~\cite{Eichten}, but not for most  energy-frontier 
topics. The latter figure implies reduction of the beam's six-dimensional (6D) phase-space volume by a factor of $\sim10^6$. This requires cooling of the longitudinal as well as the transverse degrees of freedom. 

\begin{figure}
  \includegraphics[width=\columnwidth,trim=10 60 55 110 mm]{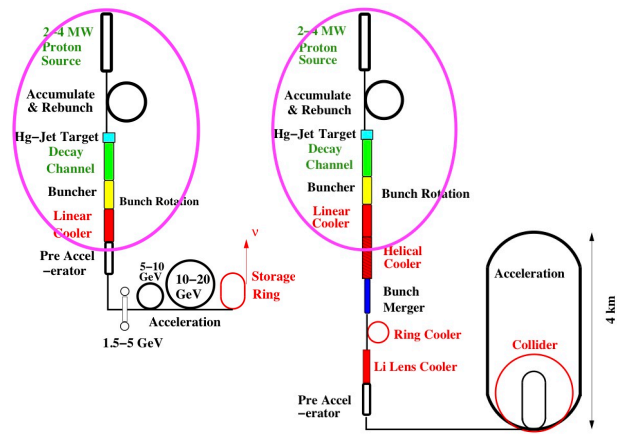}
  \caption{Comparison of (left) neutrino factory and (right) muon collider concept sketches; the front ends (circled) are similar or identical, consisting of a high-power proton driver and target, a pion decay channel, bunching and phase rotation, and a transverse cooling channel.}
  \label{fig:comparison}
\end{figure}

Muon collider R\&D is carried out under the auspices of the US Muon Accelerator Program (MAP)~\cite{MAP}. Recent muon collider designs take the neutrino factory ``front end'' as a starting point (see Fig.~\ref{fig:comparison}), allowing a straightforward staging of the two facilities. This approach makes technical sense in that a 6D cooling channel can be more effective once some transverse cooling of the beam has been accomplished. Moreover, the formation of a  bunch train in the neutrino factory front end, prior to phase rotation and cooling, reduces the individual bunch intensities, mitigating possible space-charge, wakefield, or beam-loading effects in the later stages of the cooling channel or in the acceleration chain. It also allows phase rotation and cooling using relatively efficient and economical high-frequency (200--300\,MHz) RF cavities, rather than the $<$100\,MHz required if a single bunch is to be formed and manipulated.

The staging of a neutrino factory as a first step towards construction of a muon collider is also appealing in that the high-intensity proton driver and high-power target systems can be almost identical in the two facilities. This exploits the serendipitous coincidence that a $10^{21}$-decays/year neutrino factory and a $10^{34}$-luminosity muon collider can each be driven by a 4\,MW proton beam. It also allows the needed multi-billion-dollar expenditure to be spread across multiple projects and fiscal years. It may turn out that the neutrino factory and muon collider must run in parallel in order to carry out their physics programs, rendering the neutrino factory front end  unavailable, and requiring a new front end to be built for the muon collider. Even in this scenario, the experience of building and operating a neutrino factory will prove invaluable to the muon collider project, allowing costs to be better understood and controlled, and technical risk to be significantly mitigated.

\section{Muon Collider Beam Cooling}

\begin{figure}
  \includegraphics[width=.99\columnwidth,trim=0 430 555 10 mm]{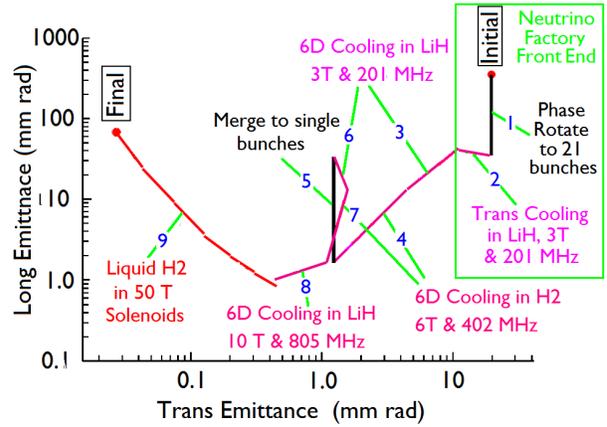}
  \caption{Trajectory in 6D phase space of a possible ``high-emittance'' muon collider cooling system (from \protect\cite{Palmer-PAC07}). Note that while the figure indicates a 50\,T final-cooling channel, simulations have shown that 40\,T solenoids are in fact sufficient to meet the HEMC specification of Table~\protect\ref{tab:params}.}
  \label{fig:Fernow-Neuffer}
\end{figure}

Table~\ref{tab:params} shows example parameters for low- (LEMC) and high-emittance muon collider (HEMC) concepts; the latter is currently the MAP baseline approach. A possible ``cooling trajectory'' from the initial captured muon-beam emittance to the final one in the HEMC approach is illustrated in Fig.~\ref{fig:Fernow-Neuffer}~\cite{Palmer-PAC07}. To put this into context, one should consider the ionization cooling equation~\cite{Neuffer}:
\begin{equation}
\frac{d\epsilon_N}{dz}=-\frac{1}{\beta^2}\left|\frac{dE_\mu}{dz}\right|\frac{\epsilon_N}{E_\mu}+
\frac{\beta_\perp(0.014\,{\rm GeV})^2}{2\beta^3E_\mu m_\mu X_0}\,,\label{eq:cool}
\end{equation}
where $\epsilon_N$ is the beam's normalized transverse emittance, $z$ the path-length traversed in an energy-absorbing medium, $E_\mu$ and $m_\mu$ the  muon energy and mass, $\beta = v/c$ the beam velocity, $\beta_\perp$ the beta function (focal length) of the magnetic lattice confining the beam, 
and $|{dE_\mu}/{dz}|$ and $X_0$ the mean rate of  muon energy loss and radiation length of the absorber medium. The demonstration of transverse ionization cooling as described by Eq.~\ref{eq:cool} is in progress  in the Muon Ionization Cooling Experiment (MICE)~\cite{MICE,OtherMICE}, under construction at the UK's Rutherford Appleton Laboratory.

One sees that  cooling to a low equilibrium emittance requires that the second (heating) term in Eq.~\ref{eq:cool} be minimized by use of a short focal length (high magnetic field or field gradient) and appropriate choice of absorber material. While ionization cooling is inherently transverse, effectively six-dimensional cooling can be obtained by coupling the transverse and longitudinal degrees of freedom via suitable dispersion in the optics of the cooling lattice, such that higher-momentum muons traverse a greater length of absorber than lower-momentum ones.  Note that the reduction of normalized emittance occurs in the absorbers, but RF cavities are also required, in order to restore the lost energy and allow iteration of the cooling process.

\begin{figure}
\includegraphics[width=\columnwidth,trim=5 1 3 1 mm]{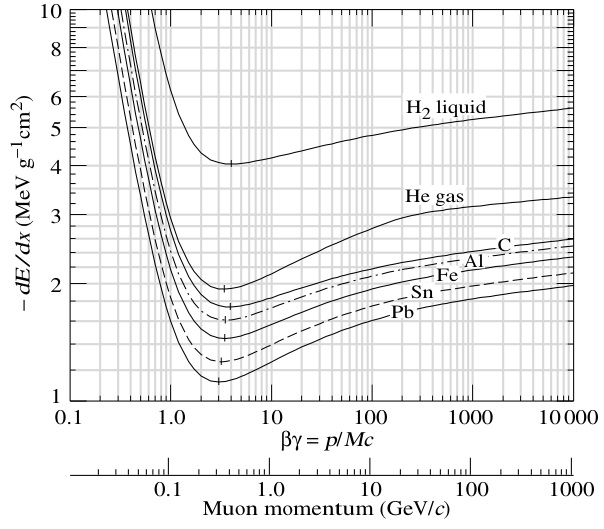}
\caption{Ionization energy-loss rate vs momentum for various media (from~\protect\cite{PDG}).}\label{fig:dEdx}
\end{figure}

\begin{figure}
\includegraphics[width=\columnwidth,trim=4 1 3 1 mm]{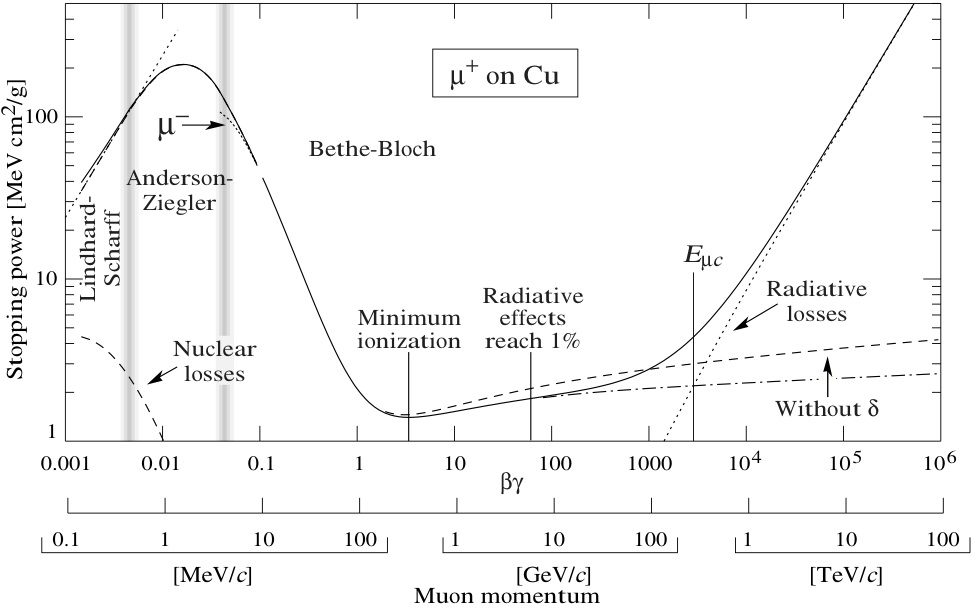}
\caption{Ionization energy-loss rate vs momentum for positive muons in copper (from~\protect\cite{PDG}).}\label{fig:dEdx-full}
\end{figure}

In the scenario of Fig.~\ref{fig:Fernow-Neuffer}, following initial transverse cooling, 6D cooling is carried out in a series of ``Guggenheim'' helical channels~\cite{Snopok} (see below), featuring liquid-hydrogen (LH$_2$) and LiH wedge absorbers, with increasingly higher RF frequencies as the emittance shrinks. The initial and 6D cooling channels operate near the ionization minimum ($\gamma\beta\approx2$), where longitudinal heating due to variation of ionization rate with energy is small (see Fig.~\ref{fig:dEdx}). 
At the ``low point'' of the trajectory in Fig.~\ref{fig:Fernow-Neuffer} [normalized  transverse and longitudinal emittance values $(\epsilon_\perp,\epsilon_{||})\approx(0.4,1)$\,mm$\cdot$rad], it becomes effective to resume transverse-only cooling, in a mode (``final cooling'') in which the longitudinal emittance is permitted to grow, and the beam energy to fall, as the transverse emittance shrinks towards the desired final value. Since, below the ionization minimum, the energy loss rate increases markedly 
as the beam energy falls (Fig.~\ref{fig:dEdx-full}),  very small transverse emittance values can be achieved with practical magnetic fields ($B\le40$\,T, achievable using high-$T_c$ superconductors at LHe temperature). 

\begin{table}
\begin{tabular}{lcc}
\hline
  & \tablehead{1}{c}{b}{LEMC}
  & \tablehead{1}{c}{b}{HEMC} \\
\hline
Avg.\ luminosity (10$^{34}$\,cm$^{-2}$s$^{-1}$) & 2.7 & 1\\
Avg.\ bending field (T) & 10 & 8\\
Rep.\ rate (Hz) & 65 & 15\\
$\beta^*$ (cm) & 0.5  & 1 \\
Muons/bunch ($10^{11}$) & 1 & 20 \\
Bunches in storage ring (each sign) & 10 & 1 \\
Norm.\ transverse emittance ($\mu$m) & 2.1 & 25\\
Norm.\ longitudinal emittance (m) & 0.35 & 0.07\\
Energy spread (\%) & 1 & 0.1 \\
\hline
\end{tabular}
\caption{Example parameters of low- (LEMC) and high-emittance (HEMC) muon collider concepts.}
\label{tab:params}
\end{table}

\section{6D Cooling}

\begin{figure}
\scalebox{.66}{\includegraphics[width=.99\columnwidth,trim=225 0 300 0mm,clip]{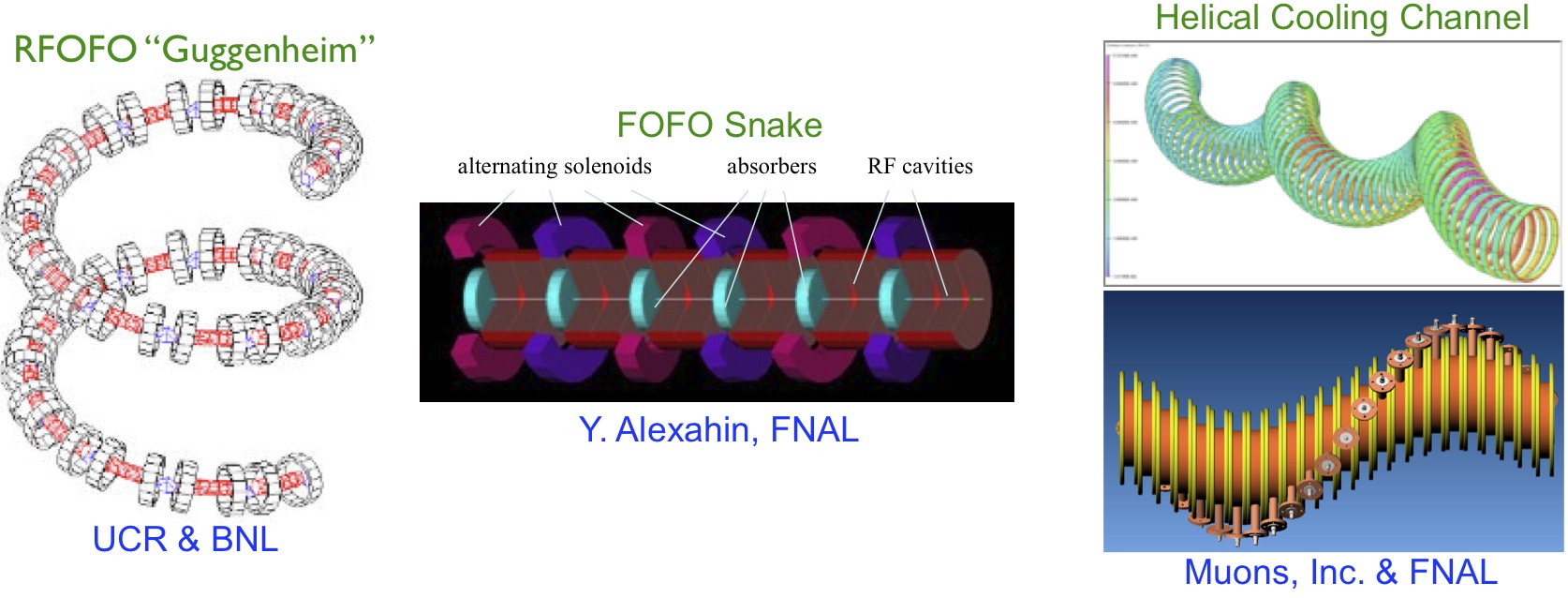}}
\qquad
\includegraphics[width=.49\columnwidth,trim=0 0 600 0mm,clip]{helices}
\qquad
\scalebox{1.2}{\includegraphics[width=.49\columnwidth,trim=560 5 0 0mm,clip]{helices}}
\caption{Three 6D cooling approaches.}\label{fig:helices}
\end{figure}

Many 6D cooling schemes have been explored since the initial muon collider study~\cite{Snowmass}. Three approaches show promise in recent simulation studies (Fig.~\ref{fig:helices}); they are dubbed the ``Guggenheim''~\cite{Snopok}, ``FOFO Snake''~\cite{Snake}, and ``Helical Cooling Channel'' (HCC)~\cite{Yonehara}. All three create dispersion by inducing helical motion of the muons using tilted or offset solenoids. Of these, the FOFO Snake channel employs the gentlest helix, phasing the solenoid tilts such that both positive and negative muons can be cooled simultaneously, on opposite phases of the RF waveform. It is thus a prime candidate for initial 6D cooling, prior to charge separation; however, its ability to cool to very small emittance appears to be limited, requiring supplementation with another cooling approach in order to reach the 
(0.4,\,1)\,mm$\cdot$rad point of Fig.~\ref{fig:Fernow-Neuffer}.

In contrast, the Guggenheim and HCC designs have both been shown capable of reaching the required emittances, but can transport only one charge at a time. By employing pressurized, hydrogen-filled RF cavities as both energy absorber and re-acceleration linac, the HCC can achieve this in a more compact configuration, but\,---\,given its array of RF feeds rotating about the helix (see Fig.~\ref{fig:helices} bottom-right), and passing between adjacent superconducting current rings\,---\,at the expense of a more complex integration challenge. 

Part of the MAP plan is to select one of these technologies for further development, based on design and prototype studies  now in progress. The selected 6D-cooling approach will then be prototyped and bench-tested. Whether a beam test will be required remains to be determined, since the principle of 6D cooling will by then have been demonstrated using wedge absorbers in MICE~\cite{MICE}.

\section{Final Cooling}

As already discussed, in the baseline muon collider scheme, 6D cooling is performed with a series of Guggenheim lattices, and final cooling with LH$_2$ absorbers in high-field solenoids.

In contrast, in the LEMC approach, HCC 6D cooling is employed, followed by a sophisticated alternate final cooling scheme featuring ``Parametric-resonance Ionization Cooling'' (PIC) and ``Reverse Emittance Exchange'' (REmEx)~\cite{Recent-Innovations}. In PIC, rather than reducing the focal length by means of increasingly strong magnetic fields, a resonance is carefully designed into the optics so as to bring about hyperbolic beam motion in transverse phase space, such that the beam divergence is constantly {\em increased}, at the expense of transverse beam area\,---\,essentially, resonant slow extraction in reverse. Since ionization cooling {\em reduces} beam divergence, this configuration can in principle cool to very small transverse emittance values. While conceptually straightforward, design of such a lattice is challenging, with possible nonlinearities, achromaticities, and aberrations all potentially important.
Despite these challenges,  progress is being made~\cite{PIC}.

A third alternative, final cooling in liquid-lithium lenses, is also receiving some attention~\cite{Balbekov,Zolkin-Cline}. Because these can carry high pulsed electric currents, they can achieve strong focusing magnetic fields, $\sim$\,300\,T/m or more, allowing cooling to sub-mm transverse emittance. Liquid-lithium lenses with such performance appear to be feasible but would require some R\&D. Hybrid schemes combining lithium lenses with high-field solenoids are also under investigation and may have beam-dynamics advantages compared to either option by itself~\cite{Balbekov}.

\section{Acceleration}
\begin{figure}
\includegraphics[width=\columnwidth,trim=0 0 0 -2mm]{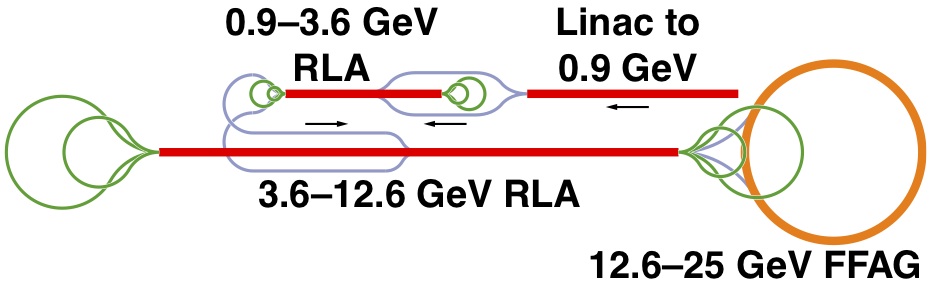}
\caption{Muon acceleration scheme from International Design Study of a Neutrino Factory~\protect\cite{IDS}.}\label{fig:IDS-accel}
\end{figure}

In any muon-beam accelerator facility, acceleration must of course be carried out very rapidly. As in the neutrino factory, initial acceleration must therefore be done in linacs. Given the low muon-beam energy at the end of the baseline final cooling section, induction linacs are the technique of choice. These should segue to the more efficient RF linacs as soon as possible, typically at muon kinetic energy of order 100\,MeV; then, at about 1\,GeV, to recirculating linacs, as in the IDS neutrino factory design~\cite{Pozimski,JINST,IDS} (see Fig.~\ref{fig:IDS-accel}).

\begin{figure}
\includegraphics[width=\columnwidth,trim=1.5 2 2 -2mm]{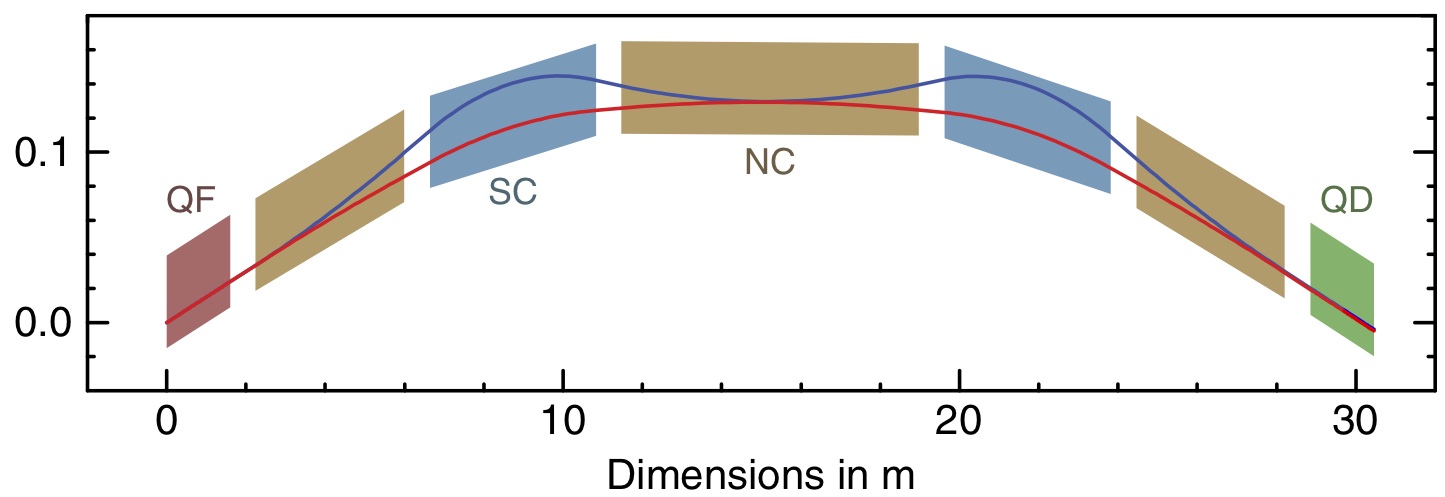}
\caption{Muon final acceleration scheme for 1.5\,TeV muon collider.}\label{fig:accel}
\end{figure}

To reach the final muon collider beam energy (perhaps 750\,GeV or more, depending on what states are found at the LHC), very rapid-cycling synchrotrons appear to be feasible and cost-effective. An example has been sketched that accelerates from 100 to 750\,GeV in two stages in a tunnel the size of the Tevatron~\cite{Summers}. The higher-energy lattice (Fig.~\ref{fig:accel}) includes 8\,T DC superconducting dipoles as well as fast-ramping normal-conducting dipoles. At injection, these bend at $-$1.8\,T (in opposition to the superconducting dipoles), then ramp  up to +1.8\,T (in accord with them). Fast-ramping quadrupoles are of course also included, with maximum gradients of 30\,T/m. To keep muon decay losses within desired limits ($<$\,10\%), this approach requires an $\sim$\,10$^3$\,Hz dipole-magnet ramp cycle. In order to reduce eddy currents to acceptable levels, the magnet laminations are formed from grain-oriented 3\%-silicon steel. Small prototypes have been tested as a proof-of-principle, and a larger-scale prototype is planned as part of MAP.

\section{Storage Ring}

\begin{figure}
\includegraphics[width=\columnwidth,trim=0 0 0 1mm]{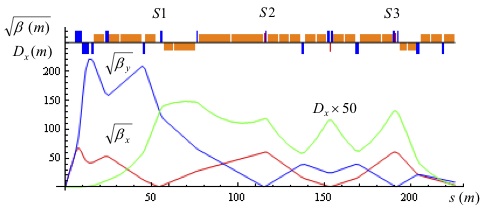}
\caption{Layout and lattice functions of muon collider interaction region in the design of Ref.~\protect\cite{Alexahin}.}\label{fig:IR}
\end{figure}

\begin{figure}
\includegraphics[width=\columnwidth,trim=0 0 0 -2mm]{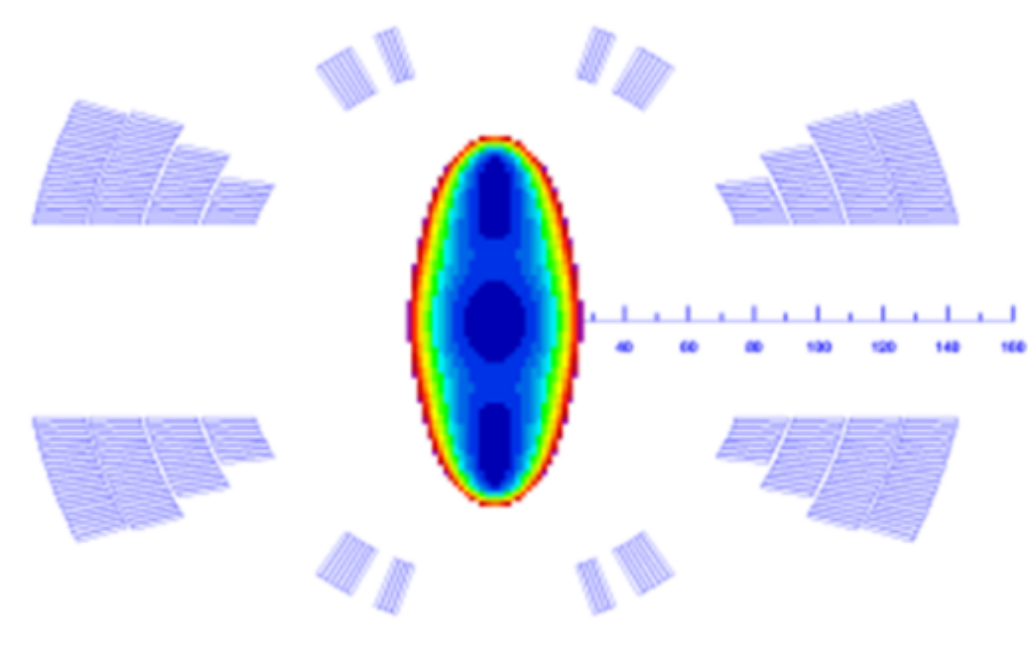}
\caption{Open-midplane dipole design: shown are a cross-section of the superconducting windings and a map of the magnetic field inside the beam pipe (from \protect\cite{Zlobin}).}\label{fig:Dipole}
\end{figure}

Given relativistic time dilation, the number of collisions per muon lifetime is directly proportional to the average bending field in the storage ring; this leads to the use of dipole fields in the neighborhood of 10\,T (Table~\ref{tab:params}), for which the number of useful beam crossings per acceleration cycle approaches 1,000. A key challenge is to avoid overheating of the superconducting magnets due to energy deposition by electrons from muon decay. A collider ring design featuring a ``dipole-first'' lattice has been explored. This would allow decay electrons swept by the dipoles closest to the interaction region (IR)  to be absorbed in small-angle tungsten cones, rather than in the low-beta quadrupoles that would normally be the nearest elements to the IR, while also creating dispersion for chromatic correction. More recently it has been superseded by a design in which the closest element is a slightly displaced quadrupole (Fig.~\ref{fig:IR}), which serves both of these purposes while also providing focusing~\cite{Alexahin}. Another important design innovation is the use of open-midplane dipoles (Fig.~\ref{fig:Dipole})~\cite{Zlobin}.

\section{Conclusions}
A program of  muon collider R\&D is in progress in the US. Its major goals for the next $\approx$\,5 years are to demonstrate the feasibility of a muon collider and provide a first cost estimate. The scheme being developed allows a natural staging from a neutrino factory to a muon collider. If the R\&D  effort is successful and such a scheme is adopted, a neutrino factory might be in operation early in the next decade, and a muon collider some years thereafter.\footnote{More detail on the MAP muon collider approach may be found in Ref.~\cite{Alexahin-RuPAC10}.}


\begin{theacknowledgments}
The author thanks his colleagues of the MAP and MICE collaborations for many stimulating and enlightening discussions. Work supported by the US Dept.\ of Energy and the National Science Foundation.
 \end{theacknowledgments}



\bibliographystyle{aipproc}   

\bibliography{sample}

\begin{thebibliography}{99}

\bibitem{Neuffer-NuFact10}
D. Neuffer, ``Muon Front-End, IDS Baseline,''
 this Workshop.

\bibitem{JINST}
M. Apollonio {\it et al.}, JINST {\bf 4} , P07001 (2009).

\bibitem{IDS}
IDS website: 
\url{https://www.ids-nf.org/}.

\bibitem{Eichten}
See e.g.\ E. Eichten, K. Lane, J. Womersley, Phys.\ Rev.\ Lett.\ {\bf 80}, 5489 (1998).

\bibitem{MAP}
MAP website: \url{http://map.fnal.gov/}.

\bibitem{Palmer-PAC07}
R. Palmer {\it et al.}, ``A Complete Scheme of Ionization Cooling for a Muon Collider,'' Proc.\ 2007 Part.\ Accel.\ Conf.\ (PAC07), Albuquerque, NM, 25--29 June 2007, paper THPMS090.

\bibitem{Neuffer}
D. Neuffer, Part.\ Accel.\ {\bf 14}, 75 (1983).

\bibitem{MICE}
Y. Torun, ``Mice Overview -- Status and Facility,'' this Workshop.

\bibitem{OtherMICE}
M. Apollonio, ``Design of the MICE Muon Beam Line,'' this Workshop;\\
M. Rayner, ``Analysis of the Performance of the MICE Muon Beam Line,'' this Workshop;\\
MICE website: \url{http://mice.iit.edu/}.

\bibitem{Snopok}
P. Snopok, G. Hanson, A. Klier, IJMPA  	
{\bf 24}, 987 (2009).

\bibitem{PDG}
K. Nakamura {\it et al.}\   [Particle Data Group], J. Phys.\ G {\bf 37}, 075021 (2010).

\bibitem{Snowmass}
R. Palmer, {\it et al.} [The $\mu^+\mu^-$ Collider Collaboration], ``Muon Muon Collider: A Feasibility Study,'' submitted to Proc.\ Snowmass96, \url{http://www.cap.bnl.gov/mumu/pubs/snowmass96.html}

\bibitem{Snake}
Y. Alexahin, ``Helical FOFO snake for 6D ionization cooling of muons,''
Proc.\ 11th Int.\ Workshop on
Neutrino Factories, Superbeams and Beta Beams (NuFact09), Chicago, IL, July 20--25, 2009, ed.\ D.\,M.\ Kaplan, M. Goodman, Z. Sullivan, AIP Conf.\ Proc.\ {\bf 1222}, 313  (2010).

\bibitem{Yonehara}
K. Yonehara, Y.\,S. Derbenev, R.\,P. Johnson, ``Helical Channel Design and Technology for Cooling of Muon Beams,'' 
Proc.\ 14th Adv.\ Accel.\ Concepts Workshop, Annapolis, MD, 13--19 June 2010, ed.\ S.\,H. Gold, G.\,S. Nusinovich, AIP Conf.\ Proc.\ {\bf 1299}, 658 (2010).

\bibitem{Recent-Innovations}
R.\,P. Johnson {\it et al.}, ``Recent Innovations in Muon Beam Cooling,'' 
Proc.\ Int.\ Workshop on Beam Cooling and Related Topics (COOL05), Eagle Ridge, Galena, IL, 18--23 Sept.\ 2005, ed.\ R.\,J. Pasquinelli, S. Nagaitsev, AIP Conf.\ Proc.\
{\bf 821}, 405 (2006).

\bibitem{PIC}
V.\,S. Morozov {\it et al.}, ``Epicyclic Twin-Helix Magnetic Structure for Parametric-Resonance Ionization Cooling,'' Proc.\ 1st Int.\ Part.\ Accel.\ Conf.\ (IPAC'10), Kyoto, Japan, 23--28 May 2010, paper MOPEA042.

\bibitem{Balbekov}
V. Balbekov, ``Final Muon Cooling with Solenoids and Li Lenses,'' 
3rd Muon Collider Design Workshop, BNL, Dec.\ 2009, 
\url{http://www.cap.bnl.gov/mumu/conf/collider-091201/}.

\bibitem{Zolkin-Cline} 
T. Zolkin, ``Design and Simulation of Final Cooling for a Muon Collider,'' 
Fourth Low-Emittance Muon Collider Workshop (LEMC2009), Fermilab, June 8--12, 2009,  \url{http://www.muonsinc.com/lemc2009/presentations/Zolkin_Presentation.ppt};\\
H.\,G. Kirk {\it et al.}, 
``6 Dimensional Muon Phase Space Cooling by Using Curved Lithium Lenses,''  Proc.\ PAC07 ({\it op cit.}), paper THPMS022.

\bibitem{Pozimski}
J. Pozimski, ``Status of Neutrino Factory R\&D Including IDR,'' this Workshop.

\bibitem{Summers}
D.\,J. Summers {\it et al.}, ``Muon Acceleration to 750 GeV in the Tevatron Tunnel for a 1.5 TeV $\mu^+\mu^-$ Collider,'' Proc.\ PAC07 ({\it op cit.}), paper THPMS082.

\bibitem{Alexahin}
Y. Alexahin {\it et al.}, ``Muon Collider Interaction Region Design,'' Proc.\ IPAC'10 ({\it op cit.}), paper TUPEB022. 

\bibitem{Zlobin}
A.\,V. Zlobin {\it et al.}, ``Magnet Designs for Muon Collider Ring and Interaction Regions,'' Proc.\ IPAC'10 ({\it op cit.}), paper MOPEB053.

\bibitem{Alexahin-RuPAC10}
Y. Alexahin,  ``Muon Collider Design Status,'' Proc.\ XXII Russian Part.\ Accel.\ Conf.\ (RuPAC-2010), Protvino, Russia, Sept.\ 28 -- Oct.\ 1, 2010, paper TUCHY01.

\end{thebibliography}

\IfFileExists{\jobname.bbl}{}
 {\typeout{}
  \typeout{******************************************}
  \typeout{** Please run ''bibtex \jobname" to optain}
  \typeout{** the bibliography and then re-run LaTeX}
  \typeout{** twice to fix the references!}
  \typeout{******************************************}
  \typeout{}
 }


\end{document}